\documentclass[pra,
                preprint,
                amsmath,
                amssymb,
                showpacs,
                superscriptaddress]{revtex4-1}
\usepackage{graphicx} 
\usepackage{dcolumn}  
\usepackage{bm}       
\usepackage{amsfonts}
\usepackage{dsfont}
\usepackage{csquotes}
\usepackage{newunicodechar}
\usepackage{ulem}
\usepackage{color}

\usepackage{float}
\usepackage{physics}
\usepackage{comment}

\begin{document}

\title{How massless are Weyl fermions in Weyl semimetals?}

\author{Amar Bharti}
\affiliation{%
Department of Physics, Indian Institute of Technology Bombay,
           Powai, Mumbai 400076, India }

\author{Misha Ivanov}     
\affiliation{%
Max-Born Institut, Max-Born Stra{\ss}e 2A, 12489 Berlin, Germany }

\author{Gopal Dixit}
\email[]{gdixit@phy.iitb.ac.in}
\affiliation{%
Department of Physics, Indian Institute of Technology Bombay, Powai, Mumbai 400076, India }
\affiliation{%
Max-Born Institut, Max-Born Stra{\ss}e 2A, 12489 Berlin, Germany }

\date{\today}

\pacs{}

\begin{abstract}
Circularly polarized light fails to generate currents in inversion-symmetric 
Weyl semimetals with degenerate Weyl nodes. While each node generates current with the 
direction depending on its chirality, the two currents  in the two degenerate nodes of opposite   
chirality cancel each other. By extension, it is also generally expected that 
the currents generated at the same Weyl node by the fields of opposite helicity should also observe mirror
symmetry and cancel. Surprisingly, here we find that this is not the case. 
The origin of this effect
lies in the nonlinear energy dispersion, 
which manifests strongly 
already very close to the Weyl nodes, where linear dispersion is expected to hold 
and the Weyl fermions are thus expected to be massless. 
A scheme based on using a trefoil field composed of a counterrotating 
fundamental and its second harmonic is proposed  to control the induced asymmetry 
at a chiral node from positive to negative, including zero. 
\end{abstract}


\maketitle 

\newpage 

Condensed matter  systems provide attractive platforms to realize exotic particles, 
originally proposed in high-energy physics. 
Weyl semimetals are one such system in which  
low-energy collective excitations are governed  by massless Weyl fermions 
which appear in pairs of opposite chirality~\cite{armitage2018weyl}. 
These  fermions exhibit  novel phenomena, such as negative magnetoresistance~\cite{parameswaran2014probing,huang2015observation,zhang2016signatures}, 
the chiral magnetic effect~\cite{vazifeh2013electromagnetic,li2016chiral,kaushik2019chiral}, the quantized circular photogalvanic effect~\cite{de2017quantized,rees2020helicity},  and the Hall effect~\cite{burkov2014anomalous,chan2016chiral,shekhar2018anomalous,li20203d}, among others~\cite{gao2020chiral,osterhoudt2019colossal,okamura2020giant,liu2020semimetals,lv2017experimental,hamara2023helicity, bharti2023role, khanna2014proximity, neufeld2023band}. 
Moreover, Weyl fermions are promising for upcoming quantum technologies at room temperature~\cite{kharzeev2019chiral,kharzeev2020quantum,chen2022room}.

Light-driven optical response
has played a pivotal role in understanding and probing 
exotic properties of Weyl semimetals~\cite{sirica2019tracking, ma2017direct,ma2019nonlinear, lv2021experimental, lv2021high, orenstein2021topology, bharti2022high}.
One such optical response is  circularly polarized light-driven selective excitations in the vicinity of the  Weyl nodes. The excitation process depends on  
the chirality of the Weyl fermions and  the helicity 
of circularly polarized light~\cite{yu2016determining}.
Helicity-driven selective excitations in  broken inversion-symmetric Weyl semimetals lead
to population asymmetry around the  Weyl nodes and  the circular photogalvanic effect:
the generation of current upon irradiation with circular light~\cite{konig2017photogalvanic,chan2017photocurrents,ma2017direct}. 
Broken inversion symmetry in Weyl semimetals is a prerequisite to ensure noncancellation of the contribution from a pair of chiral Weyl nodes.
Thus, when a measurement of coupling between the massless fermions and circularly polarized light is 
integrated over both nodes, the nonzero result arises only in the inversion-broken Weyl semimetals~\cite{ma2017direct}.  

Since this conclusion assumes perfectly massless Weyl fermions, 
i.e., a gapless system with a perfectly linear dispersion near the nodes, it welcomes a question: how 
quickly is this assumption violated as one moves away from the exact location of the node?
Note that deviations from linear dispersion imply that even for gapless nodes, the mass becomes nonzero as soon as one moves away from the degenerate point.
Can circularly polarized light with opposite helicity generate non-mirror-symmetric  excitations in inversion-symmetric Weyl semimetals 
once the nonlinearity of the band structure is taken into account, even near the Weyl nodes?
We show that the answer to the latter question is positive.

We begin with the perfectly massless Weyl fermions, where the population induced around the chiral Weyl 
node with $\chi =1$ by right circularly polarized  light is superimposable with that of $\chi = -1$ induced 
by the left circularly polarized light and vice versa. 
These populations provide the reference for  the more general case of nonlinear band dispersion. 
Once quadratic corrections to the Weyl equation 
are included, helicity-sensitive asymmetric excitations becomes nonzero and significant already at the Weyl nodes. 
That is, the excitation generated with one helicity at the $\chi=1$  node is no longer superimposable with that 
generated by the opposite helicity at the $\chi=-1$ node
and the excitations at a given node for light with opposite helicities are not mirror symmetric. 
The same result obtains for the more general  
inversion-symmetric Hamiltonian of a Weyl semimetal.
While the  induced asymmetry reduces when decreasing the light frequency, so that
the resonant excitations are located very close to the node, it still remains substantial.  
Last but not least, we devise a scheme based on two-color counterrotating circularly polarized light to 
control the helicity-sensitive asymmetric excitation. 
Our control scheme can tailor the asymmetry from positive to zero to negative.

\begin{figure}[h!]
\includegraphics[width=\linewidth]{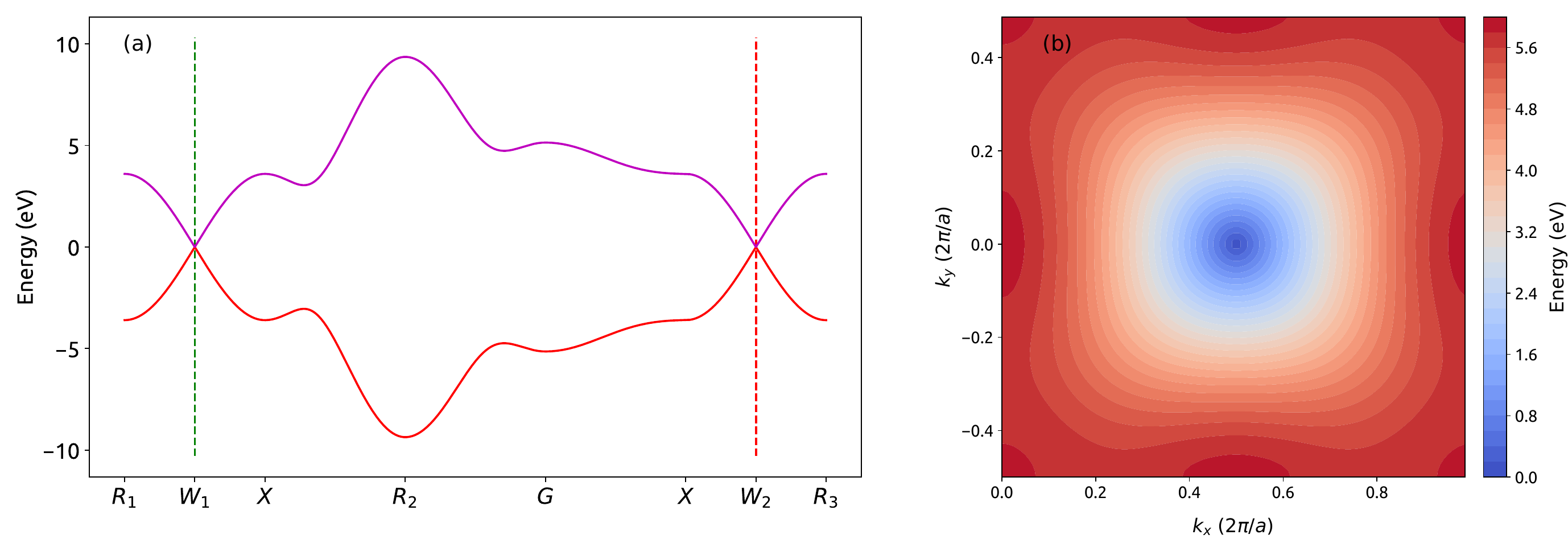}
\caption{(a) Energy dispersion along high-symmetry points of an inversion-symmetric 
Weyl  semimetal as given in Eq.~(\ref{eq:eqham}). (b) Energy contour around one of the Weyl nodes in $k_x-k_y$ plane (Weyl planes). 
The hopping parameter  is $t =1.8$ eV  and the lattice parameters are  $a = 6.28$ \AA  \ and $\beta = 0.8$. 
The lattice vectors are $a_1 = (a,-a,0), a_2 = (a,a,0)$, $a_3 =(0,0,a)$, and the
reciprocal vectors are $b_1= ( \pi/a, -\pi/a, 0),  b_2= ( \pi/a, \pi/a, 0), b_3= ( 0, 0 , 2\pi/a)$, leading 
to reduced coordinates for the high-symmetry points as follows: 
$R_1 (\pi/2a,-\pi/2a,-\pi/a)$, $X (\pi/2a,-\pi/2a,0)$, $R_2 (0,\pi/a,-\pi/a)$, $G (0,0,0)$, and  $R_3 (\pi/2a,-\pi/2a,\pi/a)$.} \label{fig:fig1}
\end{figure}

A Hamiltonian for a type-I Weyl semimetal can be written as~\cite{hou2016weyl}
\begin{equation}\label{eq:eqham}
\mathcal{H}(\mathbf{k}) = 2t_{x}\cos(k_{x} a)\sigma_{x} + 2t_{y}\cos(k_{y} a)\sigma_{y} 
+ 2t_{z}\left[\cos(k_{z} a) - \alpha - \beta \sin(k_{x} a)\sin(k_{y} a) \right] \sigma_{z},
\end{equation} 
where $t$'s are hopping parameters, $\sigma$'s are Pauli matrices, and  $\alpha$ and $\beta$ are dimensionless parameters. 
The Hamiltonian corresponds to an inversion-symmetric Weyl semimetal with broken time-reversal symmetry~\cite{hou2016weyl}. 
To make our discussion simple, we have considered 
$t_{x, y, z} = t$ and $\alpha = \beta$. 
Diagonalization of Eq.~(\ref{eq:eqham}) yields the band structure shown in Fig.~\ref{fig:fig1}(a). 
The two Weyl nodes are positioned at 
$W_{1, 2} = (\frac{\pi}{2a}, -\frac{\pi}{2a}, \pm \frac{\pi}{2a})$,  i.e., $(0.5,0,\pm0.25)$ in reduced coordinates~\cite{NoteX},  and 
are at the Fermi level.
The energy contours in their vicinity in the $k_{x} - k_{y}$ plane are isotropic [see Fig.~\ref{fig:fig1}(b)], so that
light-induced excitation should yield a symmetric population. 

Let us first focus on the linear part of the band dispersion. Expanding Eq.~(\ref{eq:eqham}) up to linear terms 
near the Weyl nodes, we find
\begin{subequations}\label{eq:Eqlin1}
\begin{align}
\mathcal{H}_{1}(\mathbf{k}) & = d_{1, x}(\mathbf{k})\sigma_{x} + d_{1, y}(\mathbf{k})\sigma_{y} + d_{1, z}(\mathbf{k})\sigma_{z} \\
\mathcal{H}_{2}(\mathbf{k}) & = d_{2, x}(\mathbf{k})\sigma_{x} + d_{2, y}(\mathbf{k})\sigma_{y} + d_{2, z}(\mathbf{k})\sigma_{z} 
\end{align}
\end{subequations}
Here, $\mathbf{k}$ denotes the deviation from the  Weyl node [for both nodes, Eqs.~(\ref{eq:Eqlin1}a) and (\ref{eq:Eqlin1}b)], 
$d_{1(2), x}(\mathbf{k}) = v \left(-k_{x} a\right), d_{1(2), y}(\mathbf{k}) = v \left(k_{y} a \right)$, and 
$d_{1(2), z}(\mathbf{k}) = v \left[-(+) \tilde{k}_{z} a \right]$, where  
$-\tilde{k}_{z}(+\tilde{k}_{z})$  is measured relative to the Weyl node 1 (2), and $v = 2t$.  
The above  Hamiltonian in Eq.~(\ref{eq:Eqlin1}) 
represents  the Weyl equation and can be written as $\mathcal{H}_w = v  ~\mathbf{k} \cdot  \bm{\sigma}$. 
As pointed above, the two Weyl 
nodes described by $\mathcal{H}_{1}(\mathbf{k})$ and $\mathcal{H}_{2}(\mathbf{k})$ are degenerate and only differ by chirality, 
which is defined   as $\chi = \textrm{sgn}(d_{x}\cdot d_{y} \times d_{z}$).
The Weyl nodes 1 and 2  have $\chi= 1$ and -1, respectively.

\begin{figure}[h!]
\includegraphics[width=\linewidth]{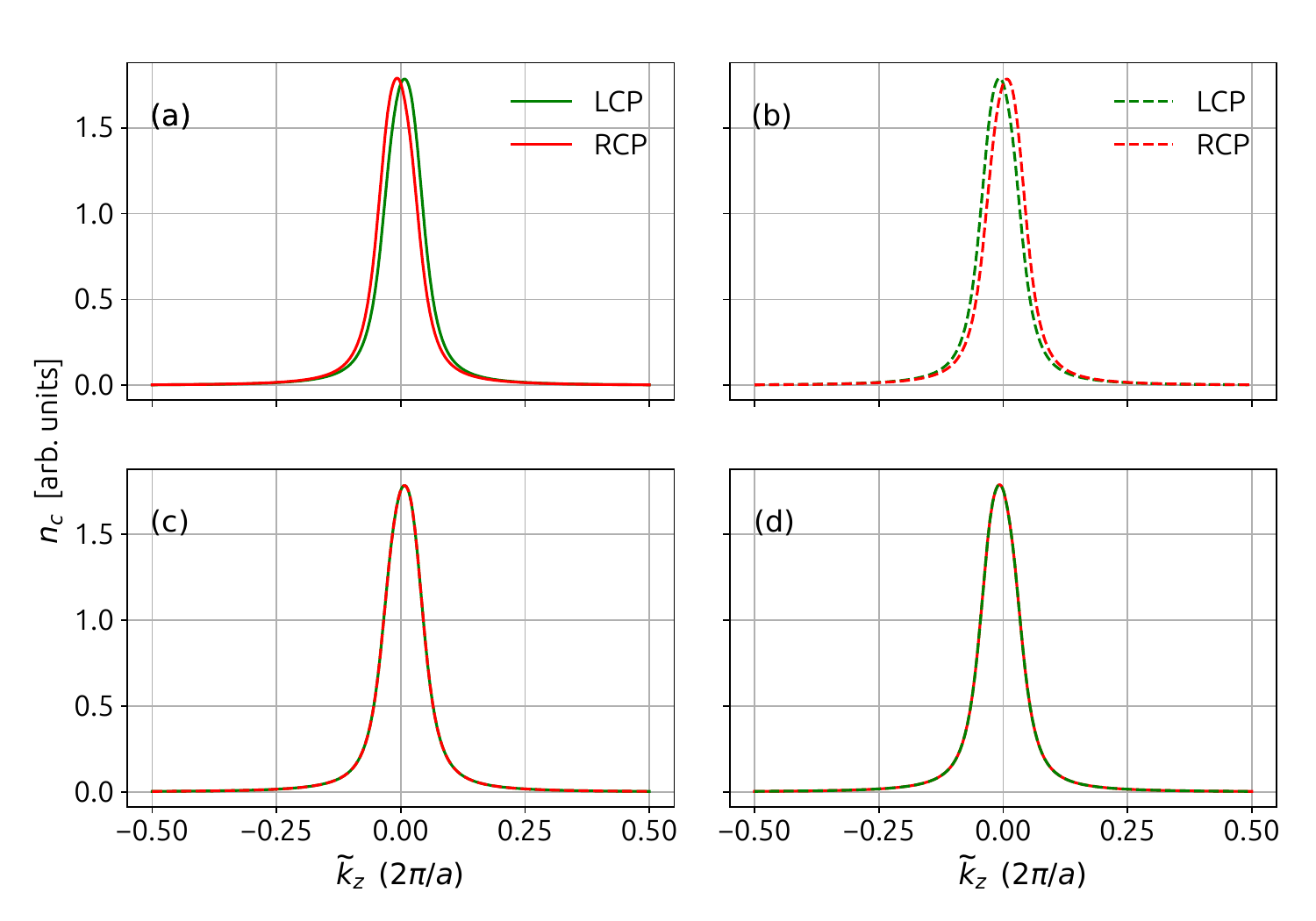}
\caption{Residual population in the conduction band ($n_{c}$) after the end of the left-handed circularly polarized (LCP)  and right-handed circularly polarized (RCP) 
llight around a Weyl node with (a)  $\chi = -1$  and (b) $\chi = 1$.  
(c) Comparison of the residual populations from a Weyl node with $\chi = -1$  due to LCP, and from a Weyl node with $\chi = 1$  due to RCP. (d) Same as (c) for a Weyl node with $\chi = -1$  due to RCP, and from a Weyl node with $\chi = 1$  due to LCP.  The Weyl nodes 
with  $\chi = -1$  and $\chi = 1$ are described by  Eq.~(\ref{eq:Eqlin1}). }\label{fig:fig2}
\end{figure}

Light-driven electronic excitation  in a Weyl semimetal is simulated using the density matrix approach within the semiconductor Bloch equations framework as discussed in Refs.~\cite{mrudul2021high, rana2022high, bharti2022high}. 
To account for the decoherence between electron and hole during the excitation process, 
a phenomenological dephasing term with 1.5 fs is introduced.  Our findings are robust against the 
dephasing term ranging from 1.5 to 10 fs.

The conduction band population  is obtained by integrating 
the density matrix in the conduction band after  the end of the laser pulse; 
the population is integrated over $k_x$ and $k_y$, and is shown along the $\tilde{k}_z$ direction, where 
$\tilde{k}_z = 0$ is the Weyl plane which contains both chiral Weyl nodes, for Eq. \eqref{eq:Eqlin1}.  
We used  $\sim$ 100 fs long circularly polarized pulses with intensity  10$^{11}$ W/cm$^2$ and  
wavelength 3.2 $\mu$m (i.e., $\omega$ = 0.39 eV); different wavelengths upto 10.6 $\mu$m 
(i.e., $\omega$ = 0.12 eV) were also studied, with the results described below.

Figure~\ref{fig:fig2} shows  
the final population around the two Weyl nodes in the conduction band after the end of the pulse, for 
$\chi = -1$ (a) and $\chi = 1$ (b) calculated for the Hamiltonians in Eqs.~(\ref{eq:Eqlin1}a) and (\ref{eq:Eqlin1}b), respectively.    
As expected, the population asymmetry is zero at $\tilde{k}_z = 0$ and is mirror symmetric with respect to 
changing either 
the light helicity or the chirality of the node. In particular, the population at $\chi = -1$ induced by 
the left circularly polarized (LCP) pulse is the same as that induced at $\chi=+1$ by 
the right circularly  polarized (RCP) pulse; see Fig.~\ref{fig:fig2}(c). 
The same is true for the population induced by the RCP at  $\chi = -1$  compared to the population induced by 
LCP near  $\chi = 1$; see Fig.~\ref{fig:fig2}(d).

\begin{figure}[h!]
\includegraphics[width=\linewidth]{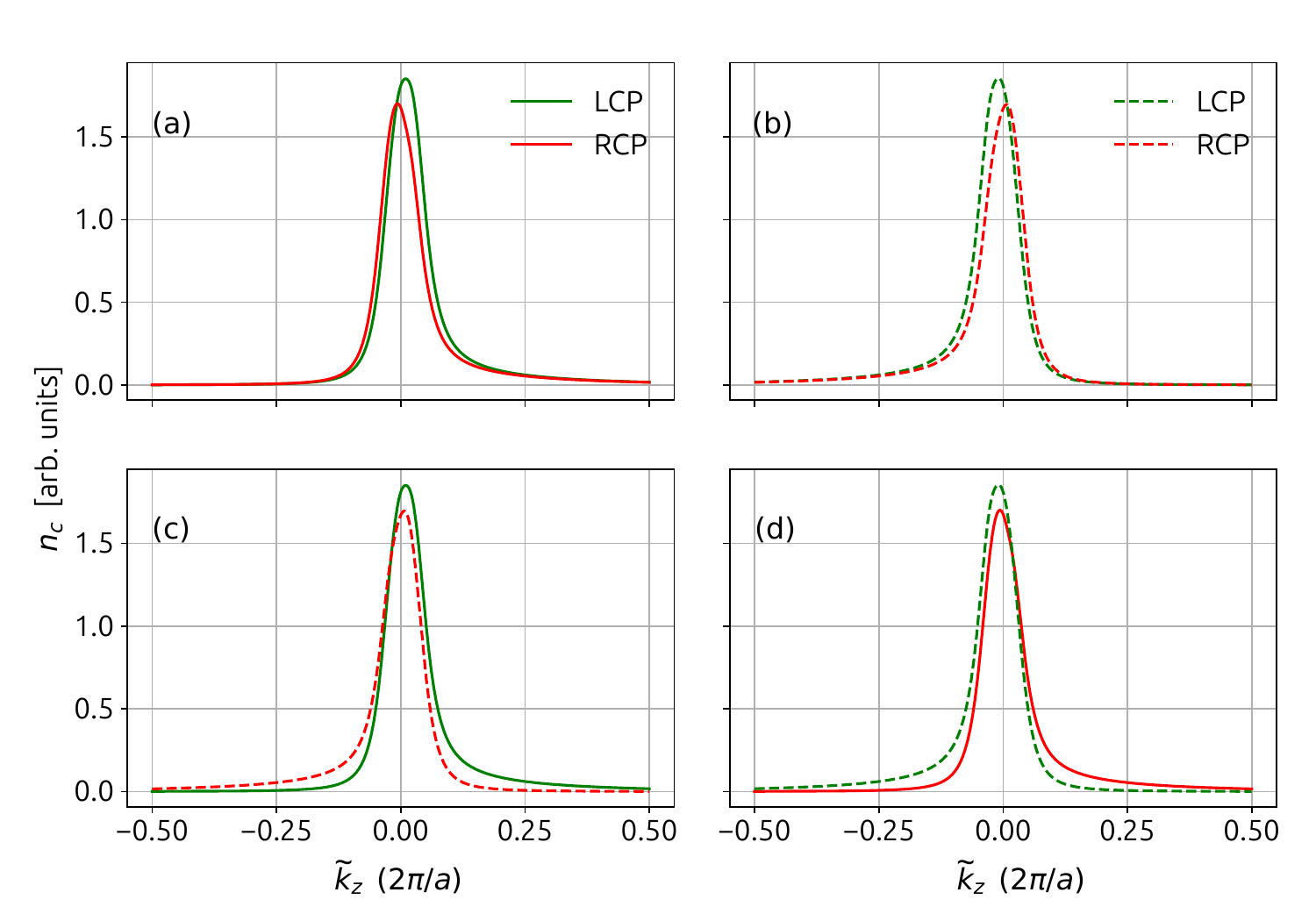}
\caption{Same as in Fig.~\ref{fig:fig2} for the Weyl nodes with $\chi = -1$  and $\chi = 1$, but now 
using the  Hamiltonian Eq.~(\ref{eq:Eqquad1}) which includes the quadratic terms.}\label{fig:fig3}
\end{figure}

Having established this reference, 
we now go beyond the linear approximation and 
expand Eq.~(\ref{eq:eqham}) to the second order, resulting in the following expression,
\begin{subequations}\label{eq:Eqquad1}
\begin{align}
 \tilde{\mathcal{H}}_{1}(\mathbf{k}) & = d_{1, x}\sigma_{x} + d_{1, y}\sigma_{y} +  \tilde{d}_{1, z}\sigma_{z},  \\
 \tilde{\mathcal{H}}_{2}(\mathbf{k}) & = d_{2, x}\sigma_{x} + d_{2, y}\sigma_{y} +  \tilde{d}_{2, z}\sigma_{z},  
\end{align}
\end{subequations}
where  $\tilde{d}_{1(2), z}(\mathbf{k}) = v \left[-(+) \tilde{k}_{z} a \right] - \tilde{v} \left[\frac{(k_{x} a)^2 + (k_{y} a)^2}{2} \right]$ with $\tilde{v} = 2t \alpha$. 
The $\tilde{d}_{z}$ component now contains additional terms quadratic in $k_{x}$ and  $k_{y}$, whereas  $d_{x}$ and $d_{y}$ remain identical in both the equations.

\begin{figure}[h!]
\includegraphics[width=\linewidth]{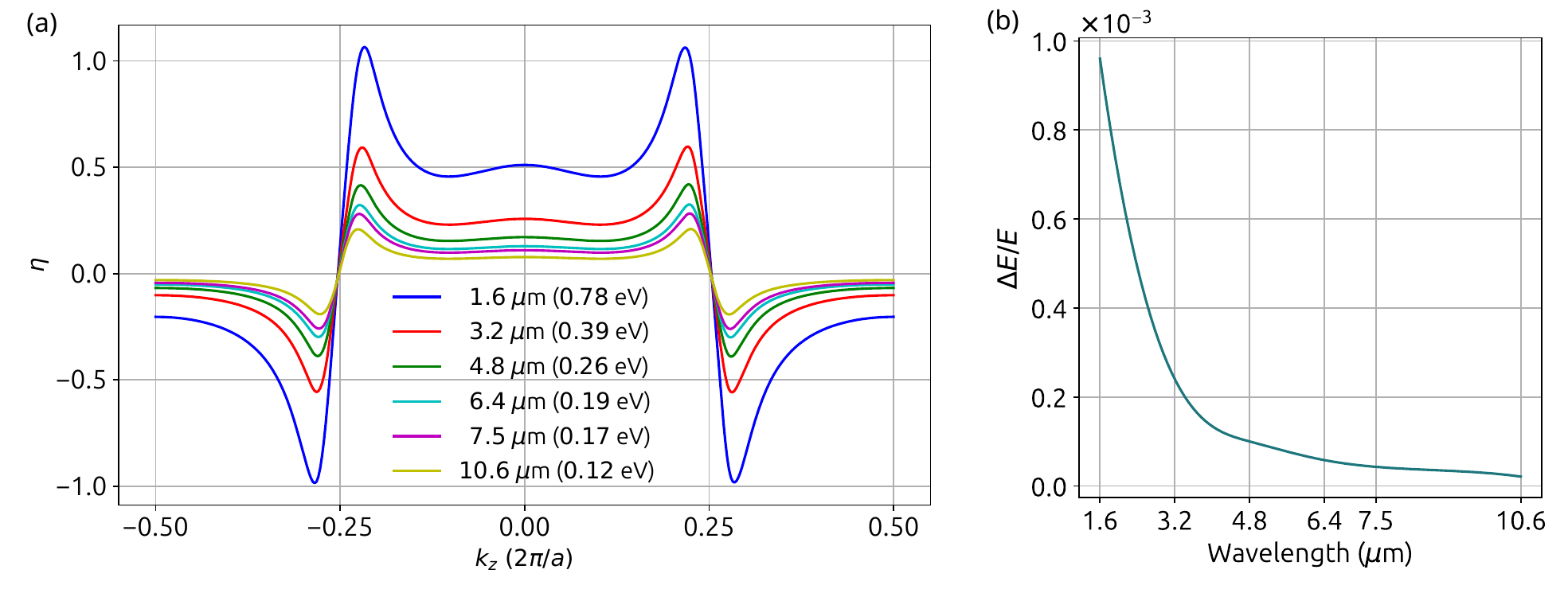}
\caption{(a) Normalized population asymmetry ($\eta$)  as a function of the wavelength of the circularly 
polarized driving pulse with intensity $5\times 10^{9}$ W/cm$^2$. (b) Nonlinear correction ($\Delta E$) 
to the energy ($E$) obtained from the linear dispersion.
The simulations use the full Hamiltonian given in Eq.~(\ref{eq:eqham}). } \label{fig:fig4}
\end{figure}

The quadratic terms affect the final population already in the immediate vicinity of the 
Weyl nodes 
as visible from 
Figs.~\ref{fig:fig3}(a) and ~\ref{fig:fig3}(b). 
The mirror symmetry upon changing the handedness of the Weyl node is, of course, preserved: the 
population near $\chi=-1$ is mirror symmetric with that near $\chi=+1$ with respect to
changing $\tilde{k}_z \rightarrow - \tilde{k}_z$. However, for a given Weyl node, the peaks of the populations induced by RCP and LCP light 
do not coincide. Similarly, the excitation induced near the $\chi=-1$ node by LCP pulse does not overlap
with the excitation induced near the $\chi=+1$ node by RCP pulse; see Fig.~\ref{fig:fig3}(c). Likewise, 
 the excitation induced near the $\chi=+1$ node by the LCP pulse does not overlap
with the excitation induced near the $\chi=-1$ node by RCP pulse; see Fig.~\ref{fig:fig3}(d).
This stands in stark contrast with Fig.~\ref{fig:fig2}.
The fact that this asymmetry, associated with
the deviations from the linear dispersion, arises in the immediate vicinity of the nodes, 
i.e., in what is supposed to be the zero-mass region, raises the question posed in the title of this Letter: 
How massless are the Weyl fermions under practical conditions of typical laser wavelengths and intensities?  

Since the deviations from the massless behavior could have come from our specific 
choice of the laser wavelength and intensity, which could have forced the electrons to explore the nonlinear 
parts of the dispersion, we will scan the laser intensity and wavelength while 
using the full Hamiltonian given in Eq.~(\ref{eq:eqham}). 
Below we shall use 
normalized population asymmetry  defined as
\begin{equation}\label{eq:eqasym}
\eta = \frac{n_{c}^{\circlearrowright} - n_{c}^{\circlearrowleft}}
{(n_{c}^{\circlearrowright} + n_{c}^{\circlearrowleft})/2},
\end{equation}
where $n_{c}^{\circlearrowright}  (n_{c}^{\circlearrowleft})$ is the final population 
due to LCP (RCP) light along $k_z$, integrated in the $k_x-k_y$ plane. 

\begin{figure}[h!]
\includegraphics[width= \linewidth]{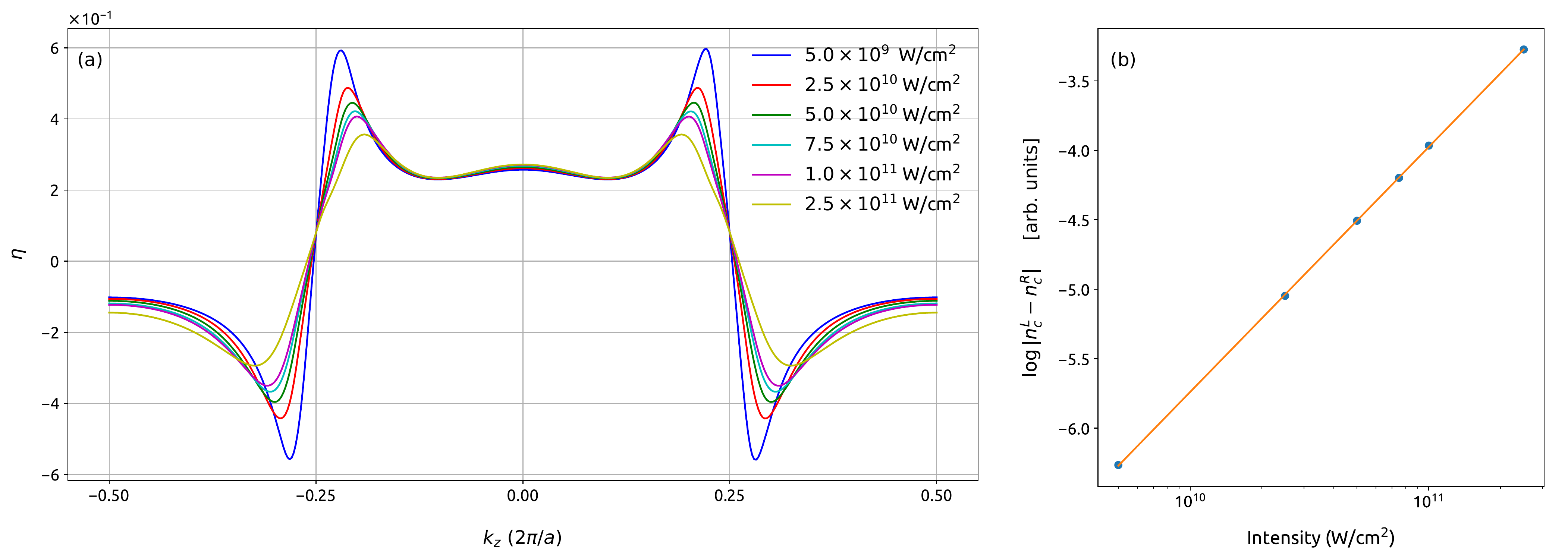}
\caption{ (a) Variations of the normalized population asymmetry ($\eta$) 
for different intensities of the  circularly polarized light. 
(b) Logarithm of the difference in the population around a given Weyl node excited by
the left- or right-handed circularly polarized light, as a function of the laser's  intensity. 
The slope of the fitted line is $0.8$, i.e., is below unity expected for linear processes.  
The driving light wavelength is  $3.2~\mu$m.  
The simulations use 
the full Hamiltonian given in  Eq.~(\ref{eq:eqham}).}
\label{fig:fig5}
\end{figure}

Figure~\ref{fig:fig4}(a) shows $\eta$ for the driving wavelengths $\lambda$ 
from $1.6~\mu$m to  $10.6~\mu$m, which allows one to access 
different parts of energy dispersion during the excitation. 
We see that for all $\lambda$ the asymmetry
$\eta$ is nonzero around  the Weyl nodes at $k_z=\pm 0.25$. While the 
asymmetry reduces with $\lambda$, even for the 
longest wavelength substantial values of $\eta$ at the levels $\sim 10$\% 
arise in the immediate vicinity of the Weyl nodes.  
We note that the deviation from linear dispersion for the wavelength studied is below 0.001 shown in 
Fig.~\ref{fig:fig4}(b), 
while the circular dichroism asymmetry induced is several orders of magnitude higher as in Fig.~\ref{fig:fig4}(a).

 \begin{figure}[]
\includegraphics[width= 0.6\linewidth]{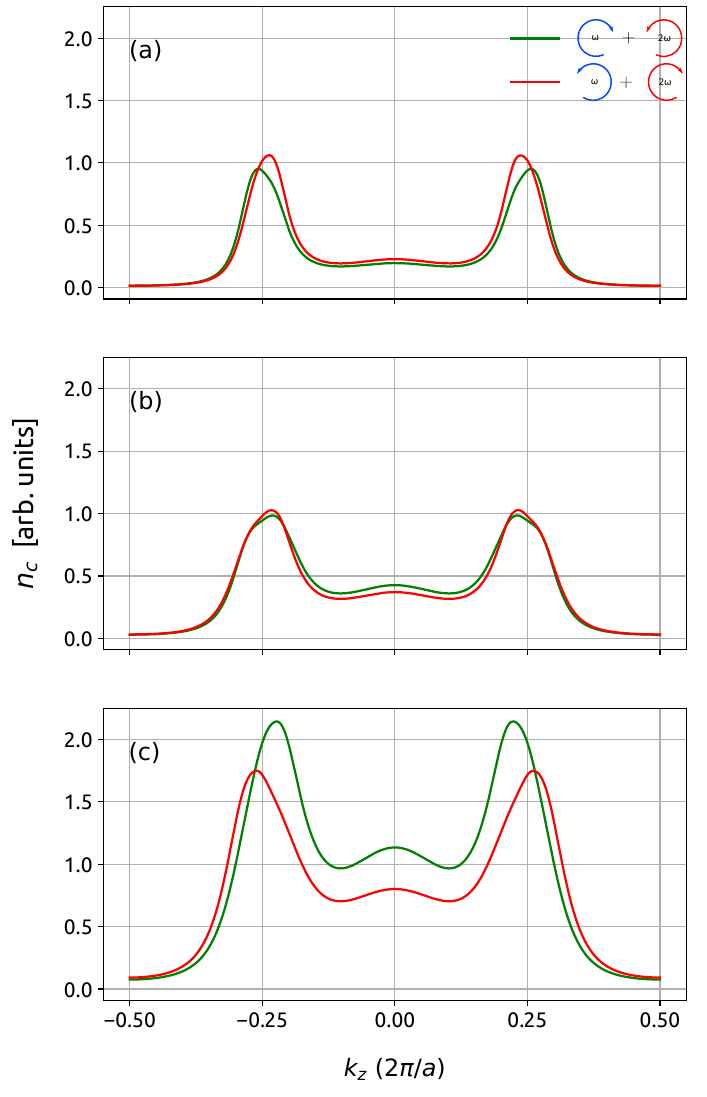}
\caption{Residual conduction-band population for different ratio  ($\mathcal{R}$)   of the  two-color $\omega-2\omega$ laser pulses:   
(a) $\mathcal{R}=0.2$, (b) $\mathcal{R}=0.5$, and (c) $\mathcal{R}=1.0$. 
Intensity and wavelength of the $\omega$ pulse are $5\times10^{10}$ W/cm$^2$ and 
$3.2~\mu$m, respectively.} \label{fig:fig6}
\end{figure}

Figure~\ref{fig:fig5}(a) shows the dependence of $\eta$  on laser intensity, for 
$\lambda=3.2 ~\mu$m.  Notably, we find that the asymmetry is nonzero 
exactly at the Weyl node, where the dispersion is linear. This is true for all laser intensities,  
with the position of the zero asymmetry moving away from the node with increasing intensity.
The another surprise is that the asymmetry decreases 
with increasing intensity, i.e., when the electron is driven to explore a wider range of the 
Brillouin zone, where the dispersion nonlinearity is stronger.  This observation is
supported by Fig.~\ref{fig:fig5}(b), which shows that the intensity dependence of the 
non-normalized asymmetry is sublinear, with the slope $0.8$.

At this point, it is natural to explore the possibilities to control the ratio of the  asymmetry induced by LCP and RCP light at a given node. 
To this end, we apply $\omega-2\omega$ counterrotating circularly polarized laser pulses with the total vector potential given by 
\begin{equation}\label{eq:eqtwocolor}
\mathbf{A}(t) = \frac{A_0 f(t)}{\sqrt{2}} \left( \left[ \cos(\omega t + \phi) + \mathcal{R} \cos(2 \omega t) \right] \hat{\mathbf{e}}_{x} \right. + \left.\left[ \sin(\omega t + \phi) - \mathcal{R} \sin(2 \omega t)\right] \hat{\mathbf{e}}_{y} \right).
\end{equation}
The ratio between the two electric fields is controlled by $\mathcal{R}$, and $\phi$ describes  the subcycle relative phase 
between the $\omega$ and $2 \omega$ pulses. 
In recent years, $\omega-2\omega$ circularly polarised  pulses have been employed to control the valley asymmetry in pristine graphene~\cite{mrudul2021light, mrudul2021controlling}.

The  population excited by $\omega-2\omega$ counterrotating pulses is shown in Fig.~\ref{fig:fig6}, with 
the fundamental wavelength $\lambda=3.2~\mu$m.
For  $\mathcal{R} = 0.2$, the RCP-LCP combination generates more excitation
than the LCP-RCP combination; see Fig.~\ref{fig:fig6}(a). 
Moreover, the peak induced  by RCP-LCP combination  leans toward  the center of the Brillouin zone.  
As  $\mathcal{R}$ changes from  0.2 to 0.5, both combinations yield almost similar population. 
However, the peaks due to LCP-RCP combination change its direction and peaked toward the center. 
The situation reverses for $\mathcal{R} = 1$ where the LCP-RCP combination generates higher excitation than the
RCP-LCP combination; see Fig.~\ref{fig:fig6}(c). 
The reason behind such behavior is the interplay of the two competing resonant processes driven by LCP and RCP light, 
which is controlled by changing  $\mathcal{R}$. 
Thus, the ratio and the behavior of the residual population can be controlled by tailoring the value of 
$\mathcal{R}$ in $\omega-2\omega$ counterrotating pulses.

In conclusion, we have demonstrated the generation of helicity-sensitive population in an inversion-symmetric Weyl semimetal, which
is not symmetric with respect to the helicity of the driving circular light. The effect is general
and persists for different wavelengths and intensities. Even for the longest wavelegnths and weakest 
intensities studied, it is triggered   by the deviations of the Weyl fermion 
mass from zero, even in the immediate vicinity of the Weyl node.  
The origin of this phenomenon is embedded in the Berry connection, which remains unaffected by any modifications in the Hamiltonian of the Weyl semimetal~\cite{sadhukhan2021role}. 
We have proposed  a way to control and manipulate  the asymmetric population using 
counterrotating bicircular light, which allow tailoring the asymmetry 
from positive to negative via nearly zero.
The asymmetric residual population  can be probed  via time- and angle-resolved photoemission spectroscopy 
in a pump-probe setup \cite{weber2021ultrafast}.  

G. D. acknowledges financial  support from SERB India  (Project No. MTR/2021/000138).   


\begin{thebibliography}{42}
\bibitem [{Not()}]{NoteX}%
\BibitemOpen
\href@noop {} {}\bibinfo {note} {The band structure and the position of the 
Weyl points obtained from Eq. (1) are periodic with a translation of reciprocal lattice vectors.}
\BibitemShut {Stop}%
\end{thebibliography}
%

\end{document}